\documentclass[a4paper,fleqn,usenatbib]{mnras}
\usepackage{newtxtext,newtxmath}
\usepackage[T1]{fontenc}
\usepackage{ae,aecompl}
\usepackage{threeparttable}
\usepackage{graphicx}
\usepackage{amsmath}
\usepackage{amssymb}

\title{Median Statistics Estimate of the Galactic Rotational Velocity}
\author[T. Camarillo et al.]{
Tia Camarillo,$^{1}$\thanks{E-mail: tiacamarillo@phys.ksu.edu}
Pauline Dredger,$^{1}$\thanks{E-mail: pmdredge@ksu.edu}
Bharat Ratra$^{1}$\thanks{E-mail: ratra@phys.ksu.edu}
\\
$^{1}$Department of Physics, Kansas State University,116 Cardwell Hall, Manhattan, KS 66506, USA
}

\date{Accepted XXX. Received YYY; in original form ZZZ}
\pubyear{2018}
\begin{document}
\label{firstpage}
\pagerange{\pageref{firstpage}--\pageref{lastpage}}
\maketitle

\begin{abstract}
We compile a complete collection of 19 recent (since 2000) measurements of $\Theta_0$, the rotational velocity of the Milky Way at $R_0$ (the radial distance of the Sun from the Galactic center). These measurements use tracers that are believed to more accurately reflect the systematic rotation of the Milky Way. Unlike other recent compilations of $\Theta_0$, our collection includes only independent measurements. We find that these 19 measurements are distributed in a mildly non-Gaussian fashion and a median statistics estimate indicates $\Theta_0 = 220\ \pm 10$ km\,s$^{-1}$ (2$\sigma$ error) as the most reliable summary, at $R_0 = 8.0 \pm 0.3$ kpc (2$\sigma$ error).
\end{abstract}

\begin{keywords}
(cosmology:) cosmological parameters -- Galaxy: fundamental parameters -- Galaxy: kinematics and dynamics -- Galaxy: structure -- methods: data analysis -- methods: statistical
\end{keywords}


\section{Introduction}

A more accurate model of the Milky Way will improve the accuracy of inter- and extra-galactic measurements. Two constants play a fundamental role in describing the current model of the Milky Way: $R_0$ (the radial distance of the Sun to the Galactic center, Sgr A*) and $\Theta_0$ (the rotational velocity of the Milky Way at $R_0$). \citet[~hereafter~C18]{camarillo2018} have recently measured $R_0$ from a carefully compiled set of independent $R_0$ data points. Earlier compilations of data used to estimate $R_0$ included non-independent measurements. C18 found that the data set they put together of 28 independent $R_0$ results published during 2011-17 was somewhat non-Gaussian and the results of their median statistics analysis are reasonably well summarized as $R_0 = 8.0 \pm 0.3$ kpc (2$\sigma$ error). After C18 appeared, J. Vall\'{e}e encouraged us to perform a similar analysis for $\Theta_0$.

There have been three recent attempts at measuring $\Theta_0$ from compilations of measurements: \citet[~hereafter~V17]{vallee2017}, \citet[~hereafter~dGB17]{dgb2017}, and \citet[~hereafter~RD18]{rajan2018}. These analyses use compilations that include non-independent measurements which can significantly affect the results and render them unreliable. In this paper we first put together a collection of 29 independent estimates of $\Theta_0$ that have been published in 2000 or later. Of these 29 measurements, 19 correspond to tracers (such as CO and H I gas clouds) that are believed to more accurately reflect the systematic rotation of the Milky Way; these are the ones we use to estimate $\Theta_0$. We find that this collection of 19 measurements is somewhat non-Gaussian so a median statistics analysis \citep{gott2001} is needed for a more reliable estimate of $\Theta_0$. Using median statistics we find  $\Theta_0 = 219.70\ ^{+\,6.67}_{-\,7.43}\ ^{+\,8.77}_{-\,10.75}$ km\,s$^{-1}$ (1$\sigma$ and 2$\sigma$ error bars) which for most purposes can be summarized as $\Theta_0 = 220 \pm 7 \pm 10$ km\,s$^{-1}$. Given the extent to which our data compilation is only mildly non-Gaussian, it is likely that undiscovered systematic errors will not significantly change these estimates and $\Theta_0 = 220 \pm 10$ km\,s$^{-1}$ (2$\sigma$ error) probably provides the most reliable estimate.

In {\S} \ref{sec:data} we discuss our compilation of recent independent $\Theta_0$ measurements and how it differs from those of V17 and dGB17. In {\S} \ref{sec:methods} we summarize the central estimates statistics and the tests of Gaussianity. We present and discuss our results {\S} \ref{sec:results}. We conclude in {\S} \ref{sec:conclusion}.

\begin{table*}
\tiny
    \caption{Independent $\Theta_0$ measurements since 2000}
    \label{table:data}
    \begin{tabular}{ccclcp{5.5cm}}
    \hline
    Radius (kpc)&$\Theta_0$ (km\,s$^{-1}$)&Rescaled $\Theta_0^\mathrm{res}$ (km\,s$^{-1}$)&Reference&Tracer Type&Notes\\
    \hline
        6.72 $\pm$ 0.39 & 203.35 $\pm$ 12.00 & 240.87 $\pm$ 20.59 & \cite{branham2014} & Young & Table 3, Hipparcos catalog, 6288 OB stars \\
        7.62 $\pm$ 0.32 & 205.00 & 214.15 $\pm$ 10.09 & \cite{bat2013} & Old & Figure 3, 4400 carbon stars  \\
        7.64 $\pm$ 0.32 & 217.00 $\pm$ 11.00 & 226.09 $\pm$ 15.63 & \cite{bob2013} & Young & Table 3, Cepehids near Sun, UCAC4 \\
        7.97 $\pm$ 0.15 & 226.80 $\pm$ 4.20 & 226.52 $\pm$ 7.69 & \cite{mcmillan2017} & Both & In Abstract, from alternative mass model \\
        7.98 $\pm$ 0.79 & 238.54 $\pm$ 11.66 & 237.94 $\pm$ 26.76 & \cite{shen2010} & Young & From Hipparcos Cepheids \\
        8.00 & 220.80 $\pm$ 13.60 & 219.70 $\pm$ 14.32 & \cite{bedin2003} & Old & From WFPC2/HST photometry on M4 globular cluster \\
        8.00 $\pm$ 0.50 & 202.70 $\pm$ 24.70 & 201.69 $\pm$ 27.95 & \cite{kalirai2004} & Old & From HST on M4 globular cluster, independent of \cite{bedin2003} \\
        8.00 $\pm$ 0.50 & 236.00 $\pm$ 15.00 & 234.82 $\pm$ 21.52 & \cite{rb2004} & Old & From VLBA proper motion around Sgr A* \\
        8.00 & 208.50 $\pm$ 20.00 & 207.46 $\pm$ 20.39 & \cite{xue2008} & Old & Averaged from Table 3 for range $7.5-8.5$ kpc \\
        8.00 & 243.50 $\pm$ 13.00 & 242.28 $\pm$ 13.93 & \cite{yuan2008} & Old & From Hipparcos K-M giants \\
        8.00 & 226.84 & 225.71 $\pm$ 4.82 & \cite{sharma2011} & Old & From comparing galaxy model to Hipparcos, Geneva-Copenhagan survey, and SDSS \\
        8.00 & 218.00 $\pm$ 10.00 & 216.91 $\pm$ 10.98 & \cite{bovyrix2013} & Old & Figure 20, 16,269 G-type dwarfs, SEGUE \\
        8.00 $\pm$ 0.40 & 234.00 $\pm$ 14.00 & 232.83 $\pm$ 18.82 & \cite{bobbaj2015} & Young & Data from spectroscopic binaries, in Results section \\
        8.00 $\pm$ 0.40 & 230.00 $\pm$ 15.00 & 228.85 $\pm$ 19.43 & \cite{bobbaj2015} & Young & Data from Calcium stars distance scale \\
        8.00 & 236.00 & 234.82 $\pm$ 5.02 & \cite{aumer2015} & Young & Page 3171, uses some APOGEE, mostly MW bar stars \\
        8.00 & 227.50 $\pm$ 5.50 & 226.36 $\pm$ 7.30 & \cite{mcgaugh2016} & Old & From CO and H I clouds, error is average of provided upper and lower error bars \\
        8.00 & 210.00 $\pm$ 10.00 & 208.95 $\pm$ 10.90 & \cite{rojas2016} & Old & Figure 8, from thin disk stars in Gaia-ESO survey \\
        8.00 $\pm$ 0.40 & 230.00 $\pm$ 12.00 & 228.85 $\pm$ 17.24 & \cite{bobbaj2016} & Young & In Abstract, from RAVE4 \\
        8.00 $\pm$ 0.20 & 231.00 $\pm$ 6.00 & 229.85 $\pm$ 9.63 & \cite{bob2017} & Young & From Gaia DR1 Cepheids \\
        8.00 $\pm$ 0.20 & 219.00 $\pm$ 8.00 & 217.91 $\pm$ 10.71 & \cite{bobbaj2017} & Young & From Gaia DR1 OB stars \\
        8.01 $\pm$ 0.44 & 202.00 $\pm$ 4.00 & 200.74 $\pm$ 12.48 & \cite{avedisova2005} & Old & From 270 star forming regions \\
        8.20 $\pm$ 0.70 & 215.00 $\pm$ 24.00 & 208.71 $\pm$ 29.67 & \cite{nikiforov2000} & Old & From 5 H I data sets \citep{nikiforov1994} and 2 CO cloud catalogs \citep{bb1993}, differs from \cite{mcgaugh2016} \\
        8.20 & 238.00 & 231.03 $\pm$ 4.93 & \cite{portail2017} & Old & Section 3.4, from red clump stars. \\
        8.24 $\pm$ 0.12 & 236.50 $\pm$ 7.20 & 228.46 $\pm$ 9.12 & \cite{rast2017} & Old & In Abstract, error from quadrature addition of error on $\Theta_0$ and range \\
        8.30 $\pm$ 0.25 & 233.00 $\pm$ 11.35 & 223.46 $\pm$ 13.66 & \cite{kupper2015} & Old & From Palomar 5 globular cluster, page 20, error is average of upper and lower provided error bars \\
        8.30 $\pm$ 0.20 & 236.00 $\pm$ 6.00 & 226.33 $\pm$ 9.29 & \cite{bob2016} & Young & In Abstract, from MWSC open-clusters catalog \\
        8.34 $\pm$ 0.16 & 240.00 $\pm$ 6.00 & 229.06 $\pm$ 8.72 & \cite{huang2016} & Old & From LAMOST/LSS-GAC and SDSS/SEGUE and SDSS-III/APOGEE, differs from \cite{bovyrix2013}; little overlap with \cite{aumer2015} \\
        8.40 $\pm$ 0.40 & 224.00 $\pm$ 12.50 & 212.27 $\pm$ 16.22 & \cite{koposov2010} & Old & From SDSS photometry and spectrometry, USN0-B astrometry, and Calar Alto telescope \\
        8.50 & 226.00 & 211.64 $\pm$ 4.52 & \cite{mb2017} & Old & Section 2.4, from galactic mass modeling \\
    \noalign{\vskip 1mm} \hline
    \end{tabular}
\end{table*}

\section{Data} \label{sec:data}

Table~\ref{table:data} lists the $\Theta_0$ data we use in our analyses here. These are from measurements published in or after 2000 and we believe this is an exhaustive list of all such independent measurements. 

In all cases the angular velocity ${\omega_0}_i = {\Theta_0}_i/{R_0}_i$ was what was measured, so we list ${\Theta_0}_i \pm \sigma_{{\Theta_0}_i}$ and ${R_0}_i \pm \sigma_{{R_0}_i}$ in this table. ${\sigma_{\Theta_0}}_i$ and/or ${\sigma_{R_0}}_i$ are not listed in Table~\ref{table:data} if these are not given in the cited reference. In C18 we measure $R_0 \pm \sigma_{R_0} = 7.96 \pm 0.17$ kpc. We use these measurements to compute the recaled
    \begin{equation} \label{eq:rescaling}
        {\Theta_0}_i^\mathrm{res} \pm \sigma_{{\Theta_0}_i}^\mathrm{res}= \frac{R_0 \Theta_i}{{R_0}_i}\left(1 \pm \sqrt{ \left(\frac{\sigma_{R_0}}{R_0}\right)^2 + \left(\frac{{\sigma_{\Theta_0}}_i}{{\Theta_0}_i}\right)^2 + \left(\frac{{\sigma_{R_0}}_i}{{R_0}_i}\right)^2}\right)
    \end{equation}
and list these in column 3 of Table~\ref{table:data}.\footnote{More properly one would use the rescaled angular velocities in the analysis and then convert the resulting angular velocity central value to a linear velocity central value. However, the uncertainty on $R_0$ is small and so results from the two different approaches will only differ slightly.} 

It has been known for quite a while now that $\Theta_0$ measured using different tracers differ (\citealp{roman1950,roman1952,avedisova2005,yuan2008}; dGB17, and references therein). We categorize the measurements listed in Table~\ref{table:data} into either Old or Young (one publication uses both types of tracers and is listed as Both). Old tracers include CO and H I gas clouds and are thought to better reflect the systematic rotation of the Galaxy while Young tracers such as Cepheids are believed to have velocities that are contaminated by "peculiar" motions. In Table~\ref{table:data} we have $18+1$ Old measurements and $10+1$ Young ones.

Unlike the measurements listed in Table~\ref{table:data}, the collections compiled by V17 and dGB17 include non-independent data points. In their analyses dGB17 and RD18 consider different subsets of data, based on tracer type and/or year of publication, but like V17 they also do not study a compilation of independent measurements. This lack of independence can bias results. Here we have invested significant effort in compiling a collection of independent $\Theta_0$ measurements published during 2000-2017. 

\section{Methods of Analysis} \label{sec:methods}

It is expected that a large enough data set of $N$ independent measurements will follow a Gaussian distribution, however it is not unheard of for an astronomical parameter to not obey a Gaussian distribution.\footnote{Perhaps the most famous example is the Hubble constant \citep{chen2003,chenratra2011a}. For other examples in astronomy, cosmology, and physics see \cite{farooq2013,crandall2015,farooq2017,bailey2017,zhang2017}, and references therein. Significant effort is devoted to testing for intrinsic non-Gaussianity in physical systems \citep[e.g.][]{park2001,planck2016}, as opposed to measurement induced non-Gaussianity, since Gaussianity is usually assumed in parameter estimation \citep[e.g.][]{samushia2007,chenratra2011b,ooba2017}.} Here we study the data compilation of Table~\ref{table:data} (and two subcompilations) to examine if it is non-Gaussian or not. If it is non-Gaussian this could be caused by improperly estimated errors. 

To estimate the Gaussianity of a data collection we need to use a central estimate of the data. We consider two here, the median central estimate and the weighted mean central estimate.

Median statistics does not use information of the error on a measurement at all and the true median of a data set can be found independent of any of the individual measurements errors. The estimated median will have a larger uncertainty than that of a weighted central estimate statistic that makes use of error information. We use the median statistics technique developed by \cite{gott2001}. The median is defined as the value with 50\% probability of finding another value above and below it. \cite{gott2001} show that for a dataset of $i=1,2,...,N$ independent values, $\Theta_i$, the probability of the median being between $\Theta_i$ and $\Theta_{i+1}$ is given by the binomial distribution
\begin{equation}
    \label{eq:gott}
    {P}={\frac{{{2}^{-N}}{N!}}{{i!}(N-1)!}}.
\end{equation}
The $1\sigma$ error about the median is then defined by the range about it such that $68.27\%$ of the probability is included. This can be extended to finding the $2\sigma$ error about the median, where instead $95.45\%$ of the probability would be enclosed. We refer to the median of the Galactic rotational velocity at $R_0$ as $\Theta_0^{\mathrm{med}}$.

The weighted mean comes with the benefit of additional information in the errors, at the potential expense of including inaccurate uncertainties \citep{podariu2001}. The weighted mean of the Galactic rotational velocity is
\begin{equation}
    \Theta_0^{\mathrm{wm}} = \frac{\sum_{i=1}^{N}{\Theta_0}_i/{\sigma_{\Theta_0}}_i^{2}}{\sum_{i=1}^{N}1/{\sigma_{\Theta_0}}_i^{2}},
\end{equation}
where ${\Theta_0}_i$ and ${\sigma_{\Theta_0}}_i$ are the rotational velocities and errors. The weighted mean standard deviation is
\begin{equation}
    \sigma_{\Theta_0}^\mathrm{wm}=\frac{1}{\sqrt{\sum_{i=1}^{N}1/{\sigma_{\Theta_0}}_i^{2}}}.
\end{equation}

The next step in analyzing the data is to construct error distributions of the data based on the chosen central estimate. For a central estimate $\Theta_0^\mathrm{CE}$ independent of the data ${\Theta_0}_i$, the number of standard deviations that each value deviates from the central estimate is 
\begin{equation}
    {N_{\sigma_{\Theta_0}}}_i=\frac{{\Theta_0}_i-\Theta_0^\mathrm{CE}}{\sqrt{\ \mathrm{Var\ }({\Theta_0}_i-\Theta_0^\mathrm{CE})}}
\end{equation}
where $\mathrm{Var\ }({\Theta_0}_i-\Theta_0^\mathrm{CE})$ is the variance between the independent measurement, ${\Theta_0}_i$ and the central estimate, $\Theta_0^\mathrm{CE}$.
    
For median statistics when the central estimate is assumed to not be directly correlated with the data itself we have
\begin{equation} \label{NsigM}
    N_{{\sigma_{\Theta_0}}_i}^\mathrm{med}=\frac{{\Theta_0}_i-\Theta_0^\mathrm{med}}{\sqrt{\left(\sigma_{{\Theta_0}_i}^\mathrm{}\right)^2+\left(\sigma_{{\Theta_0}_i}^\mathrm{med}\right)^2}}.
\end{equation}
For Gaussianly distributed measurements and the weighted mean central estimate estimated from the data (and so correlated with the data) we instead have (see the Appendix of C18)\footnote{An analogous equation for median statistics, for the case when the median is estimated from the data and so is correlated with the data, is not yet known.}
\begin{equation} \label{NsigWM}
    N_{{\sigma_{\Theta_0}}_i}^\mathrm{wm}=\frac{{\Theta_0}_i-\Theta_0^\mathrm{wm}}{\sqrt{\left(\sigma_{{\Theta_0}_i}^\mathrm{}\right)^2-\left(\sigma_{{\Theta_0}_i}^\mathrm{wm}\right)^2}}.
\end{equation}
The two error distributions, $N_{\sigma_{\Theta_0}}^\mathrm{med}$ and $N_{\sigma_{\Theta_0}}^\mathrm{wm}$, can be analyzed with a non-parametric Kolmogorov-Smirnov (KS) test.

We compare these error distributions to a few standard functional forms (\citealp[e.g.][]{crandallratra2015}; C18). The four probability distribution functions (PDFs) we consider here are the standard Gaussian PDF, the Cauchy (Lorentzian) PDF, the Student's $t$ PDF, and the Laplacian (double exponential) PDF. The KS test allows us to quantify the level of deviation from Gaussianity of the error distributions by using the outputs of this test, the $D$-statistic and the $p$-value. For a $95\%$ confidence level in the probability that we cannot reject a specific PDF as describing the data, we look for two requirements for a data set of 29 measurements: $D\to 0$ but $\leq 0.246$, and $p\to 1$. For the Old tracers (19 measurements) we require almost the same: $D\to 0$ but $\leq 0.301$, and $p\to 1$, and for Young tracers (11 measurements) we require $D\to 0$ but $\leq 0.391$, and $p\to 1$.\footnote{See C18 and Appendix 3 of \cite{PRE} for more detailed discussion of the outputs of the KS tests and the critical values for below which $D$ must fall.}

The standard Gaussian PDF is
\begin{equation} \label{eq:gaussian}
    P(|\textbf{X}|)=\frac{1}{\sqrt{2\pi}}\exp{(-|\textbf{X}|^{2}/2)},
\end{equation}
and is characterized by $1\sigma$ ($2\sigma$), or $68.27\%$ ($95.45\%$), of the data falling within $|X|\leq1$ ($|X|\leq2$). 

The Cauchy PDF has higher probability in the tails of the curve, and $1\sigma$ ($2\sigma$), or $68.27\%$ ($95.45\%$), of the data falls within $|X|\leq1.8$ ($|X|\leq14$). The Cauchy PDF is
\begin{equation} \label{eq:cauchy}
    P(|\textbf{X}|)=\frac{1}{\pi}\frac{1}{1+|\textbf{X}|^{2}}.
\end{equation}
    
The Student's $t$ distribution also has widened tails, and involves an additional parameter $n$. A Student's $t$ of $n=1$ is the Cauchy PDF, and as $n\to \infty$, it becomes the Gaussian. The PDF is 
\begin{equation} \label{eq:studentst}
    P(|\textbf{X}|)=\frac{\Gamma[(n+1)/2]}{\sqrt{\pi n}\Gamma(n/2)}\frac{1}{(1+|\textbf{X}|^{2}/n)^{(n+1)/2}}
\end{equation}
and the limits for $1\sigma$ ($2\sigma$), or $68.27\%$ ($95.45\%$), of the data varies with $n$. Here $\Gamma$ is the gamma function and the addition of $n$ (positive integer parameter) decreases the total degrees of freedom by one.

The last PDF we consider is the Laplacian
\begin{equation} \label{eq:laplace}
    P(|\textbf{X}|)=\frac{1}{2}\exp{(-|\textbf{X}|)}.
\end{equation}
It is characterized by $1\sigma$ ($2\sigma$), or $68.27\%$ ($95.45\%$), of the data falling within $|X|\leq1.2$ ($|X|\leq3.1$). This results in a more sharply peaked distribution than either a Cauchy or a Student's $t$. 

In eqs. (\ref{eq:gaussian}), (\ref{eq:cauchy}), (\ref{eq:studentst}), and (\ref{eq:laplace}) the value $|\textbf{X}|=|\Theta_0/S|$. $S$ is a scale factor width for each distribution ($S=1$ for a Gaussian PDF represents the standard normal distribution). We allow $S$ to vary in small increments of $0.001$ from $S=0\ \mathrm{to}\ 5$, and for the Student's $t$ we do this for every value of $n$ from $n=2\ \mathrm{to}\ 100$. The KS test, being non-parametric, makes no assumptions about the data.

\section{Results} \label{sec:results}
    
We provide in Table~\ref{table:CEs} the central estimate statistics for the data listed in column 3 of Table~\ref{table:data}. In Table~\ref{table:CEs}, column 2 shows the median (with $1\sigma$ and $2\sigma$ error ranges) and weighted mean results for all 29 values. Column 3 shows the results of only analyzing the 18 Old tracer references, plus the mixed tracer type of \cite{mcmillan2017}. Column 4 shows the results of the 10 Young tracer types, plus the mixed tracer type as well.

\begin{table*}  \tiny \centering
    \caption{Central estimates of rescaled $\Theta_0$ data (in km\,s$^{-1}$)}
    \label{table:CEs}
    \begin{tabular}{lccc}
    \hline
    Statistic & All Tracers & Old Tracers & Young Tracers\\
    \hline
        Median & $226.35\ ^{+\,2.12}_{-\,2.89}\ ^{+\,2.50}_{-\,8.44}$ & $219.70\ ^{+\,6.67}_{-\,7.43}\ ^{+\,8.77}_{-\,10.75}$ & $228.85\ ^{+\,3.98}_{-\,2.33}\ ^{+\,9.09}_{-\,2.76}$ \\[0.15cm]
        $1\sigma$ range & $223.46 - 228.46$ & $212.27 - 226.36$ & $226.52 - 232.83$ \\
        $2\sigma$ range & $217.91 - 228.85$ & $208.95 - 228.46$ & $226.09 - 237.94$ \\ \hline
        Weighted Mean & $224.36 \pm 1.67$ & $222.01 \pm 1.99$ & $230.05 \pm 3.09$ \\[0.1cm]
        $1\sigma$ range & $222.69 - 226.03$ & $220.02 - 224.00$ & $226.95 - 233.14$ \\
    \noalign{\vskip 1mm} \hline
    \end{tabular}
\end{table*}

\begin{table*} \scriptsize \centering
    \caption{$N_\sigma$ KS test results for rescaled $\Theta_0$}
    \label{table:ks}
    \begin{threeparttable}
    \begin{tabular}{@{\extracolsep{2pt}}llcclcc} \hline
    &\multicolumn{3}{c}{$N_{\sigma_{\Theta_0}}^\mathrm{med}$}&\multicolumn{3}{c}{$N_{\sigma_{\Theta_0}}^\mathrm{wm}$}\\ \\
    \cline{2-4} \cline{5-7}\\
    Type&PDF&$p$\tnote{a}&$S$\tnote{b}&PDF&$p$\tnote{a}&$S$\tnote{b}\\
    \hline
        All  &  Gaussian & 0.49 & 0.75 & Gaussian & 0.20 & 0.73 \\
             &  Cauchy & 0.70 & 0.46 &  Cauchy & 0.34 & 0.45 \\
             &  $n=2$ Student's $t$ & 0.59 & 0.59 &  $n=2$ Student's $t$ & 0.26 & 0.58 \\
             &  Laplace & 0.62 & 0.69 &  Laplace & 0.29 & 0.68 \\ \hline
             
        Old  &  Gaussian & 0.83\tnote{c} & 1.08 & Gaussian & 0.78 & 1.26 \\
             &  Cauchy & 0.71 & 0.70 &  Cauchy & 0.81 & 0.89 \\
             &  $n=36$ Student's $t$ & 0.83\tnote{d} & 1.07 &  $n=2$ Student's $t$ & 0.80 & 1.06 \\
             &  Laplace & 0.76 & 1.04 &  Laplace & 0.81 & 1.27 \\ \hline
             
        Young&  Gaussian & 0.99 & 0.35 & Gaussian & 0.99 & 0.38 \\
             &  Cauchy & 0.99 & 0.18 &  Cauchy & 0.99 & 0.20 \\
             &  $n=2$ Student's $t$ & 0.99 & 0.23 &  $n=2$ Student's $t$ & 0.99 & 0.26 \\
             &  Laplace & 0.99 & 0.27 &  Laplace & 0.99 & 0.30 \\
        \noalign{\vskip 1mm} \hline
    \end{tabular}
    \begin{tablenotes}\footnotesize
    \item [a] The probability ($p$-value) that the input data doesn't not come from the PDF.
    \item [b] The scale factor $S$ that maximizes $p$.
    \item [c] More precisely, $p=0.82817$.
    \item [d] More precisely, $p=0.82811$.
    \end{tablenotes}
    \end{threeparttable}
\end{table*}

While Table~\ref{table:ks} shows the highest probabilities for Young tracer types, with all probabilities $p\geq 0.99$, the scale factors for all these PDFs are very non-Gaussian with all of them having $1\sigma$ ranges requiring $|X|\leq 0.5$. The All tracers compilation is also fairly non-Gaussian.

For the Old tracers collection with the median as the central estimate, $p=0.83$ while $S=1.08$ for the Gaussian PDF, indicating not unreasonable consistency with Gaussianity. This is also supported by the weighted mean result for the Gaussian PDF. Together these results indicate that the weighted mean summary for $\Theta_0$ is less appropriate than our median statistics one of $\Theta_0 = 219.70\ ^{+\,6.67}_{-\,7.43}\ ^{+\,8.77}_{-\,10.75}$ km\,s$^{-1}$ (1$\sigma$ and 2$\sigma$ errors), which for most purposes can be taken to be $\Theta_0 = 220 \pm 7 \pm 10$ km\,s$^{-1}$. In summary, for practical purposes, we find at 1$\sigma$:

    $\Theta_0 = 220 \pm 7\ \mathrm{km}\ \mathrm{s}^{-1}$    
    
    $R_0 = 7.96 \pm 0.17\ \mathrm{kpc}$
    
    $\omega_0 = \Theta_0 / R_0 = 27.6 \pm 1.1\ \mathrm{km}\ \mathrm{s}^{-1}\ \mathrm{kpc}^{-1}$\\
where the angular speed $\omega_0$ error is determined by adding the fractional uncertainties of $\Theta_0$ and $R_0$ in quadrature.

Table 1 of V17 lists 28 measurements of $\Theta_0$ from mid-2012 to 2017. V17 arrives at a $\Theta_0$ close to $230$ km s$^{-1}$: median value $\Theta_0^\mathrm{med} = 232$ km s$^{-1}$, weighted mean value $\Theta_0^\mathrm{wm} = 228 \pm 2$ km s$^{-1}$, and an arithmetic mean value $\Theta_0^\mathrm{mean} = 229 \pm 3$ km s$^{-1}$. He recommends the set of Galactic constants:

    $\Theta_0 = 230 \pm 3\ \mathrm{km}\ \mathrm{s}^{-1}$    
    
    $R_0 = 8.0 \pm 0.2\ \mathrm{kpc}$
    
    $\omega_0 = \Theta_0 / R_0 = 29 \pm 1\ \mathrm{km}\ \mathrm{s}^{-1}\ \mathrm{kpc}^{-1}$ \\
We emphasize that several of the V17 Table 1 data are repeats of prior publications, big offenders being masers, OB stars, and Cepheids. Less than half of V17 Table 1 measurements are included in our list of independent measurements. V17 also does not distinguish between Old and Young tracer measurements of $\Theta_0$. These are probably why the V17 $\Theta_0$ differs from our estimate.

dGB17 on the other hand do note that Old tracers provide a better estimate of $\Theta_0$ and their recommended set of Galactic constants are (when their statistical and systematic errors are added in quadrature):

    $\Theta_0 = 225 \pm 10\ \mathrm{km}\ \mathrm{s}^{-1}$    
    
    $R_0 = 8.3 \pm 0.4\ \mathrm{kpc}$
    
    $\omega_0 = \Theta_0 / R_0 = 27.1 \pm 1.8\ \mathrm{km}\ \mathrm{s}^{-1}\ \mathrm{kpc}^{-1}$. \\
While their Old tracers compilation includes non-independent data points, dGB17 add on rather large undiscovered systematic errors and so their results are not inconsistent with our results. We note, in particular, as described in C18, that their estimate of $R_0$ is based on a very small set of data points (that are also not all independent). We emphasize that from our analysis of the Gaussianity of our $R_0$ and $\Theta_0$ compilations, here and in C18, we do not see strong evidence for large undiscovered systematic errors that dGB17 advocate for.

RD18 use 139 Galactic rotation speed values, 137 of which are from the online database of dGB17. Included are a number of non-independent measurements. For both median and weighted mean statistics they use the error distribution form of eq.~(\ref{NsigM}) and analyze the full collection of data as well as various subsets. RD18 were the first to realize that the dGB17 $\Theta_0$ data (and subsets) was non-Gaussian, but as they didn't discard non-independent measurements (as we have done) they found the data to be more non-Gaussian than we do. From a median statistics analysis of the full data set they recommend:

    $\Theta_0 = 219.65\ \mathrm{km}\ \mathrm{s}^{-1}$
    
    $R_0 = 8.3\ \mathrm{kpc}$
    
    $\omega_0 = \Theta_0 / R_0 = 26.46\ \mathrm{km}\ \mathrm{s}^{-1}\ \mathrm{kpc}^{-1}$. \\
They do not derive an $R_0$ value, instead they use that estimated by dGB17. They also do not estimate an error for $\Theta_0$.
    
\section{Conclusion} \label{sec:conclusion}

The data listed in Table~\ref{table:data} is the first compilation of independent $\Theta_0$ measurements published during 2000-2017. Given the mild non-Gaussianity of the Old tracer measurements, we favor a median statistics value of $\Theta_0 = 219.70\ ^{+\,6.67}_{-\,7.43}\ ^{+\,8.77}_{-\,10.75}$ km\,s$^{-1}$ (1$\sigma$ and 2$\sigma$ errors). For most purposes this can be summarized as $\Theta_0 = 220\pm 7 \pm 10$ km\,s$^{-1}$. Given that the measured non-Gaussianity is mild, we believe most current $\Theta_0$ error bars are reasonable and that at present there is no strong evidence for large undiscovered systematic errors. In summary our recommended set of Galactic constants, with $1\sigma$ error bars,

    $\Theta_0 = 220 \pm 7\ \mathrm{km}\ \mathrm{s}^{-1}$    
    
    $R_0 = 7.96 \pm 0.17\ \mathrm{kpc}$
    
    $\omega_0 = \Theta_0 / R_0 = 27.6 \pm 1.1\ \mathrm{km}\ \mathrm{s}^{-1}\ \mathrm{kpc}^{-1}$ \\
are probably the most reliable.

\section*{Acknowledgements}

We thank A. Quillen and J. Vall\'{e}e. This research was supported in part by DOE grant DE-SC0011840.


\bibliographystyle{mnras}


\bsp	
\label{lastpage}
\end{document}